\documentclass[aps,prl,twocolumn,showpacs,amssymb,%
floatfix,superscriptaddress]{revtex4}
\usepackage[english]{babel}
\usepackage{amsmath}
\usepackage{amssymb}
\usepackage{epsfig}
\usepackage{graphicx}
\usepackage{longtable}
\usepackage{soul}
\usepackage{epsf}
\usepackage{dcolumn}
\usepackage{bm}
\usepackage{hyperref}
\usepackage{latexsym}
\usepackage{color}
\usepackage[english]{babel}
\usepackage{bm}
\usepackage{url}
\usepackage{amsfonts}
\usepackage{amsthm}
\usepackage{graphicx}
\usepackage[latin1]{inputenc}

\def\sfrac#1#2{{\textstyle{#1\over #2}}}

\begin{document}

\title{The Evolution of Complex Networks: A New Framework}
\author{Guido Caldarelli} \affiliation{IMT Institute for Advanced
  Studies, Piazza San Ponziano 6, 55100 Lucca, Italy}
\affiliation{Istituto dei Sistemi Complessi, Consiglio Nazionale delle
  Ricerche, Dip. Fisica Universit\`a "Sapienza", P.le A. Moro 2, 00185
  Rome, Italy} \affiliation{London Institute of Mathematical Sciences,
  35a South St, Mayfair London W1K 2XF, UK} \affiliation{Linkalab,
  Complex Systems Computational Laboratory, 09129 Cagliari, Italy}

\author{Alessandro Chessa} \affiliation{Istituto dei Sistemi
  Complessi, Consiglio Nazionale delle Ricerche, Dip. Fisica
  Universit\`a "Sapienza", P.le A. Moro 2, 00185 Rome, Italy}
\affiliation{Linkalab, Complex Systems Computational Laboratory, 09129
  Cagliari, Italy}\author{Irene Crimaldi} \affiliation{IMT Institute
  for Advanced Studies, Piazza San Ponziano 6, 55100 Lucca, Italy}
\author{Fabio Pammolli} \affiliation{IMT Institute for Advanced
  Studies, Piazza San Ponziano 6, 55100 Lucca, Italy}

\begin{abstract}
We introduce a new framework for the analysis of the dynamics of
networks, based on randomly reinforced urn (RRU) processes, in which
the weight of the edges is determined by a reinforcement mechanism. We
rigorously explain the empirical evidence that in many real networks
there is a subset of ``dominant edges'' that control a major share of
the total weight of the network.  Furthermore, we introduce a new
statistical procedure to study the evolution of networks over time,
assessing if a given instance of the nework is taken at its steady
state or not. Our results are quite general, since they are not based
on a particular probability distribution or functional form of the
weights.  We test our model in the context of the International Trade
Network, showing the existence of a core of dominant links and
determining its size.
\end{abstract}


\maketitle

A large number of real systems in different domains, such as physics
\cite{AB}, economics \cite{castri,W,T}, computer science \cite{Fcube},
social science \cite{newmanbook}, transportation \cite{batty} and
others, can be efficiently described by a network structure, where the
nodes are the system entities and the links represent the relations
between them \cite{mybook}.  In comparison to that, relatively few
models have been presented in order to explain the onset of
scale-invariance in statistical distributions of degree and other
topological properties (as betweenness, clustering and assortativity).
In this paper we present a new model of network growth and evolution
based on the randomly reinforced urn (RRU) processes theory. The model
maps the weights of a particular edge with the number of balls of a
particular color which are added in the urn. Our model is particularly
suitable for dense and weighted networks, a situation often
problematic both for modeling and for randomization. Due to the
analytical properties of this treatment, one can define a statistical
procedure for investigating the dominance of one set of edges (colors)
vis \`a vis the others. Importantly enough, our procedure allows to
determine if a particular instance of a dynamical network is taken at
the steady state of network evolution or not.

The model builds on a recent kind of randomly reinforced urn (RRU)
processes \cite{AMS, BCPR,BCPRdom,C,MF} so that the probability of picking 
an edge (color) {\em depends} on its weight. At each time-step
(the time is beaten by the drawings) the picked edge (color) brings a
{\em random} weight (number of added balls) and at the next time step
the probability of picking a certain edge (color) is proportional, not
simply to the number of drawings of that edge (color), but to the
total weight already allocated to that edge (total number of added
balls of that color): a sort of {\em weighted preferential
  attachment}.

If we consider a network with $N$ vertices and $L$ edges (directed or
not, we typically consider complete graphs), then this dynamics
defines a weighted adjacency matrix ${\bf W}_u$ for every time-step
$u$, where the generic element $w_{uij}$ is the total weight on the
edge $i,j$ until time-step $u$ (i.e. the total number of added balls
of color $i,j$ until time-step $u$). Hereafter we indicate the various
edges by the index $\ell$ (with $\ell \in [1,L]$).  Similarly we
define a matrix ${\bf K}_u$ whose elements $k_{u\ell}=[{\bf
    K}_u]_{\ell}$ represents the total number of drawings of edge
$\ell$ until time-step $u$.

 More specifically, the dynamics of the network is the following.  We
 start at time $u=1$, by picking an edge $\ell^*=i^*,j^*$ according to
 following rule: every edge $\ell$ can be picked with an initial
 probability $Z_{0\ell}=a_{\ell}/\sum_{\ell=1}^{L} a_\ell$, where the
 parameters $a_\ell$ are strictly positive. (The actual value of these
 parameters plays no role in the asymptotic results and the
 statistical tools we will present in the sequel). After that a random
 weight $W_{1\ell^*}\geq 0$ is added to the chosen edge $\ell^*$. We
 do not pay particular attention to the specific form of these
 weights, provided that the weights are independent positive random
 variables, which are uniformly bounded by a constant. We finally pick
 a new edge according to the probability distribution given by
\begin{equation}
Z_{u\ell^*}=
\frac{a_{\ell^*}+\sum_{n=1}^{u}W_{n\ell^*}X_{n\ell^*}}
{\sum_{\ell=1}^L a_\ell+\sum_{\ell=1} ^L\sum_{n=1}^u W_{n\ell}X_{n\ell}}
\end{equation}
where $X_{n\ell}=1$ if at the $n$th time-step the edge $\ell$ was
chosen and it is defined equal to zero otherwise.  In other words we
define (akin to the preferential attachment idea) a probability of
edge-extraction that takes into account the previous growth of the
network.  We can write
\begin{eqnarray}
\left[{\bf K}_{u}\right]_{\ell}&=&\sum_{n=1}^{u}X_{n\ell} \nonumber \\
\left[{\bf W}_{u}\right]_{\ell}&=&\sum_{n=1}^{u}W_{n\ell}X_{n\ell}
\end{eqnarray}


Our model is related to weighted-network modeling, since it is
described, not only by binary adjacency matrices, but also by the
sequence $({\bf K}_u)$, which counts the number of times each edge is
picked, and the sequence $({\bf W}_u)$, which records the total weight
of each edge.

Given a subset ${\cal D}$ of the $L$ edges, we suppose that, for every 
time-step $u$, 
\begin{eqnarray}
E\left[W_{u\ell^*}\right]&=&\mu^*>0\; \forall \ell^*\in {\cal D}\,, \nonumber \\
E\left[W_{u\ell}\right]&=&\mu_\ell<\mu^* \; \forall \ell\notin {\cal D}
\end{eqnarray} 
and $Var[W_{u\ell}]=\sigma^2_\ell\in(0,+\infty)$.  If the set ${\cal
  D}$ coincides with the $L$ edges, the above conditions imply that
the weights have the same mean value for all edges. Conversely, when
the number of elements in the set ${\cal D}$ is lower than $L$ the
weights associated to the edges in $\cal D$ ``dominate in mean'' on
those associated to the others. (Note that a typical case of the first
type holds when every weight $W_{u\ell}$ is equal to $1$, i.e. the
classical preferential attachment.) Our analysis covers both these
cases.

As $u\to +\infty$, the probability $Z_{u\ell}$ of choosing the edge
$\ell$ converges almost surely (a.s.) to zero when $\ell\notin{\cal
  D}$; while it converges a.s. to a random variable $Z_{\ell^*}$ with
values in $]0,1]$ a.s. when $\ell=\ell^*\in{\cal D}$ and
    $\sum_{\ell^*\in{\cal D}} Z_{\ell^*}=1$ \cite{BCPR,BCPRdom}.
    Therefore the notion of ``dominant edges'' could provide a
    formalization of the empirical evidence that many real networks
    are rather sparse.  This means that with respect to all the
    possible edges, a club of edges collects the mayor fraction of the
    total weight of the network.  More precisely, it has been proved
    that, as the number of time-steps $u$ grows, the total weight on
    the dominant edges grows according to
\begin{equation}
\frac{\sum_{\ell\in {\cal D}}[{\bf W}_u]_\ell}{u}
=\frac{\sum_{\ell\in {\cal D}}\sum_{n=1}^{u}W_{n\ell}X_{n\ell}}{u}
\stackrel{a.s.}\longrightarrow \mu^*;
\end{equation}
while the same limit for the dominated edges is zero, i.e.
\begin{equation}
\frac{\sum_{\ell\notin {\cal D}}[{\bf W}_u]_\ell}{u}
=\frac{\sum_{\ell\notin {\cal D}}\sum_{n=1}^{u}W_{n\ell}X_{n\ell}}{u}
\stackrel{a.s.}\longrightarrow 0\,.
\end{equation}
Moreover, for a dominant edge $\ell^*$, the total weight associated
to that edge normalized by the total weight of the network (assumed to
be non zero) converges a.s. to the previous random variable
$Z_{\ell^*}$ according to
\begin{equation}
\frac{[{\bf W}_u]_{\ell^*}}{\sum_{\ell=1}^L [{\bf W}_u]_{\ell}}
=
\frac{\sum_{n=1}^{u} W_{n\ell^*}X_{n\ell^*}}
{\sum_{\ell=1}^L\sum_{n=1}^u W_{n\ell}X_{n\ell}}
\stackrel{a.s.}\sim Z_{u\ell^*}
\stackrel{a.s.}\longrightarrow Z_{\ell^*}
\end{equation} 
and the number of extractions of $\ell^*$ divided by the total number of
extractions converges a.s. to the same random variable, that is
\begin{equation}
\frac{[{\bf
K}_u]_{\ell^*}}{u}=\overline{X}_{u\ell^*}=\frac{\sum_{n=1}^u
X_{n\ell^*}}{u} \stackrel{a.s.}\longrightarrow Z_{\ell^*}\,.
\end{equation}

The corresponding limits for dominated edges are both equal to
zero. In particular, we have
$u^{1-\lambda}Z_{u\ell}\stackrel{a.s.}\longrightarrow 0$ for
$\ell\notin{\cal D}$ and each $\lambda\in(\overline\lambda,1)$ where
$\overline\lambda=\max_{\ell\notin {\cal D}}\mu_\ell/\mu^*$.

\begin{figure}
\begin{centering}
\includegraphics[width=0.95\columnwidth]{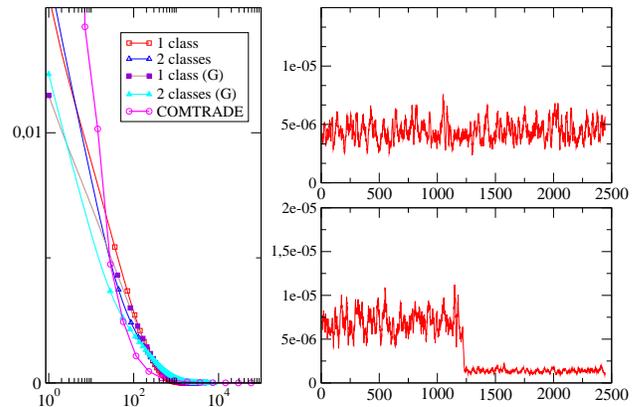}
\par\end{centering}
\caption{(color online) We performed some numerical simulations of the
  model (with $L=2500$) by preassigning both no (1 class) and one
  dominant set (2 classes). On the left we plot the frequency
  distribution of the weights in the network, for uniform and
  truncated Gaussian (G) choice of the distribution of $W$
  (for a comparison we plot also the weights distribution of the
  COMTRADE data). 
  On the right we plot the histogram of the number of drawings of each edge
  with no dominant set (up) and with the set $[1,1225]$ as the dominant
  set (below).}
\label{fig-sim1}
\end{figure}

Based on the above limit relations and some asymptotic results,
analytically proved in \cite{BCPR, BCPRdom}, we have developed a
statistical test for the class $\cal D$. In particular, we can test
the hypothesis of a given subset becoming the class of dominant edges
during the evolution of the network.  Similarly, it is possibile to
test if a particular instance of a given network has a weight
distribution that already evolved into its stationary state or not.

We assume as a null hypothesis that the ``dominant set'' ${\cal D}$
coincides with a certain subset of edges ${\cal D}^*$ with
$\mbox{card}({\cal D}^*)\geq 2$ and consider a certain level
$(1-\alpha)$ (typically $\alpha=5\%,10\%$). Then we fix $\ell^* \in
{\cal D^*}$ and compare the quantity (defined in the sequel)
\begin{equation}
\frac{|C^*_{u\ell^*}|}{\sqrt{U_{u\ell^*}}}
\end{equation}
with the quantile $q_\alpha$ of the standard normal distribution
${\cal N}(0,1)$ of order $(1-\alpha/2)$ (that is $q_\alpha$ is the
number such that ${\cal N}(0,1)(q_{\alpha},+\infty)=\frac{\alpha}{2}$
and $q_\alpha=1.96$ for $\alpha=5\%$ and $q_\alpha=1.645$ for
$\alpha=10\%$).  If the computed quantity is greater than $q_\alpha$,
then we reject the null hypothesis at the (approximate) level
$(1-\alpha)$; otherwise we can not reject it. The random variable
$U_{u\ell^*}$ is defined as
\begin{equation}
U_{u\ell^*}\!=\! \sfrac{\!\overline{X}_{u\ell^*}
  \bigl\{(1-\overline{X}_{u\ell^*})^2\widehat{\sigma}_{u\ell^*}^2+
\overline{X}_{u\ell^*} 
\sum_{\ell\in {\cal D}^*,\ell \neq \ell^*}\overline{X}_{u\ell}\,
\widehat{\sigma}_{u\ell}^2\!\bigr\}}
{(\widehat{\mu}^*_u)^2
\,\bigl(\sum_{\ell\in {\cal D}^*}\overline{X}_{u\ell}\bigr)^4}
\end{equation}
where 
$\overline{X}_{u\ell}=\sum_{n=1}^u X_{n\ell}/u$ and
$\widehat{\mu}^*_u$ is an estimate of the mean value $\mu^*$ and
$\widehat{\sigma}_{u\ell}^2$ is an estimate of the variance
$\sigma_\ell^2$, i.e.
\begin{equation}
\begin{split}
&\widehat{\mu}^*_u=\sfrac{1}{\text{card}({\cal D}^*)}
\sum_{\ell\in {\cal D}^*}\left(\sfrac{\sum_{n=1}^u W_{n\ell}X_{n\ell}}
{\sum_{n=1}^u X_{n\ell}}\right),\\
&\widehat{\sigma}_{u\ell}^2=
\sfrac{\sum_{n=1}^u W_{n\ell}^2 X_{n\ell}}{\sum_{n=1}^u X_{n\ell}}
-
\left(\sfrac{\sum_{n=1}^u W_{n\ell}X_{n\ell}}
{\sum_{n=1}^u X_{n\ell}}\right)^2\,.
\end{split}
\end{equation}
Further the random variable $C^*_{u\ell^*}$ is defined as 
\begin{equation}
C_{u\ell^*}^*=\sqrt{u}(\overline{X}_{u\ell^*}^*-Z_{u\ell^*}^*)\,,
\end{equation}
where
\begin{equation}
\begin{split}
&Z_{u\ell^*}^*=
\sfrac{1+\sum_{n=1}^u W_{n\ell^*}X_{n\ell^*}}{\mbox{card}({\cal D}^*)
+\sum_{\ell\in{\cal D}^*}\sum_{n=1}^u W_{n\ell}X_{n\ell}},
\\
&{\overline X}_{u\ell^*}^*=\sfrac{\sum_{n=1}^u X_{n\ell^*}}
{1+\sum_{\ell\in {\cal D}^*}\sum_{n=1}^uX_{n\ell}}\,.
\end{split}
\end{equation}
 
Simulations have shown that, if we perform the above test taking
${\cal D}^*$ exactly equal to the preassigned dominant set, then the
percentage of indexes $\ell^*$ for which the test gives the rejection
of the hypothesis is very low ($=2.28\%$ for $\alpha=10\%$ and
$0.82\%$ for $\alpha=5\%$). From now on we will call this percentage
the ``rejection percentage''. If we consider a different ${\cal D}^*$
with the same cardinality of the real dominant set, the rejection
percentage increases (even if we change a single element): the more
${\cal D}^*$ and the real dominat set are different, the higher the
rejection percentage is (we got values up to $93\%$ for $\alpha=10\%$
and $85\%$ for $\alpha=5\%$). However, we observed that the power of
this test decreases with the decreasing of the cardinality of ${\cal
  D}^*$. For instance, it is not able to reject the null hypothesis
when ${\cal D}^*$ is strictly contained in the real dominant set.  As
a solution to this problem, we add to the previous test another
statistical test obtained by replacing the random variable
$U_{u\ell^*}$ by
\begin{equation} 
\sfrac{\!\overline{X}_{u\ell^*}\!
\bigl\{(1-\overline{X}_{u\ell^*})^2\!\widehat{\sigma}_{u\ell^*}^2\!+\!
\overline{X}_{u\ell^*} 
\sum_{\ell\in {\cal D}^*,\ell \neq \ell^*}\!\overline{X}_{u\ell}
\widehat{\sigma}_{u\ell}^2\!\bigr\}
\!\bigl(\!\sum_{\ell\in{\cal D}^*}\!\overline{X}_{u\ell}\!\bigr)^2\!}
{(\widehat{\mu}^*_u)^2}.
\end{equation}
This second test works very well for ${\cal D}^*$ with small
cardinality (the rejection percentage goes from $80\%$ to $100\%$.)

In sum, based on these two statistical tests, we have introduced a
{\em statistical procedure to study the dominant set of a network and
  predict if a certain edge distribution will disappear in the steady
  state of the graph evolution or not}.

As an application and a test, we consider the international trade
network (ITN), also known in complex network literature as the
world-trade web \cite{SB}. ITN is defined as the network of
import-export relationships between world countries in a given period
(usually a year). Many efforts have been devoted to analyze the
structure and the dynamics of the ITN from an empirical and
theoretical modeling perspective (see, for instance,
\cite{GL-2004,HMR,BMSKM-2008,FRS-2009,RS,GCC,GL-2005,H}.  However,
existing contributions are not able to rigorously explain the evidence
that there exists a ``club of a few rich countries'' \cite{BMSKM-2008}
that control a major share of the trade network.  This issue of
``rich-club'' detection is particularly important also from a
theoretical point of view. Rich club property (i.e. the proportion of
vertices whose degree is larger than a certain value that are also
connected each other) can be defined in a proper way only for sparse
networks \cite{colizza}, while no consensus exists for the case of
dense networks \cite{gin} as ITN.  In particular, for dense networks
it is particularly difficult to define a reference or null case,
against which one can measure the specific features of the real
system.  Our model allows a natural description of this case and it
also allows for a rigorous analysis of the stability of the
statistical distributions.  In the context of the ITN, we assume that
the nodes represent the countries and the edges represent the trade
between them. With regard to the weights\cite{PNAS}, there are
different possibilities. The most natural choice is to define the
weight of a certain edge $\ell=i,j$ in terms of the value of the flow
from $i$ to $j$.

\begin{figure}
\begin{centering}
\includegraphics[width=0.95\columnwidth]{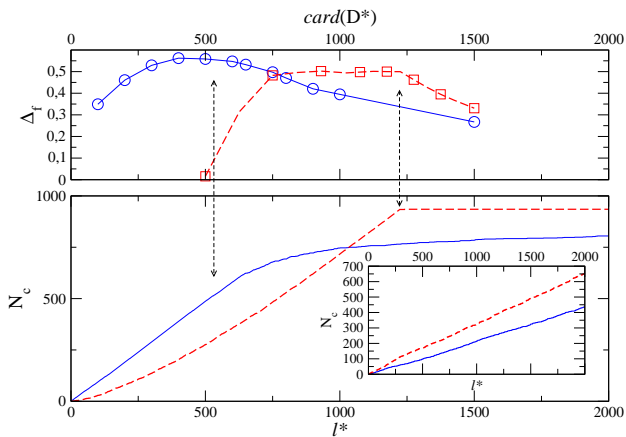}
\par\end{centering}
\caption{(color online) In the lower panel we checked the number of
  no-rejections for the COMTRADE data and the simulated data of an
  urn with colored balls in the case of uniform distributions. For
  both cases we considered a ${\cal D}^*$ of size $2000$ and ordered
  the edges/colors in descending order according to the edge 
  weight/number of balls values. We then executed the test considering
  $\ell^*$ running from the highest to the lowest value and
  accumulating the number of no-rejections in the $y$-axis.  After a
  constant no-rejection rate the COMTRADE data (blue line) start
  bending, signaling the presence of a core subset of dominant edges.
  The same happens for the urn with colored balls (red line) but with
  a much more sharp turning point, exactly in correspondence of the
  dominant ${\cal D}^*$ size of $1225$, known a priori.  In the inset the
  same procedure has been performed for a random ${\cal D}^*$ for the
  two corresponding cases.  In the upper panel, we calculated the
  difference between the rejection percentage for the ordered and the
  random case and discovered a maximum where the two curves start
  bending.}
\label{fig-comtrade}
\end{figure}

As a real case data example we consider here the data of trades
between nations in the years 1948-2000 as it is possible to
reconstruct from COMTRADE data \cite{COMTRADE}.  We computed for each
year and for each couple of countries $(A,B)$ the total exports (when
present) from $A$ to $B$. The ordered couple $(A,B)$ is an edge
(color) while the edge weight for a certain year represents an
extraction of that edge (color) where the number of added balls is the
amount of dollars for the total exports for that edge in that year.
For the COMTRADE data we don't know in advance the ``dominant edges''
set but we can leverage from the statistical test previously defined
to extract at least a core subset of it.  In order to get this core
subset we fixed ${\cal D}^*$ of size $2000$ and performed the first
test for ${\cal D}^*$ picking up $\ell^*$ in descending order starting
from the largest edge weight.  If we then plot the number of
no-rejections along the whole set of $\ell^*$ in ${\cal D}^*$, we find
that for the ordered case the number of no-rejections grows linearly
with constant slope close to $1$ but at a certain point starts bending
(see Fig. \ref{fig-comtrade}). After this bending it saturates and
reaches a plateau where the $\ell^*$ will always give a rejection.
Remarkably we found an ``optimal'' size of ${\cal D}^*$ for which the
difference of the rejection percentage for the ordered edges and the
random case is maximal, revealing that the set of top ranking edges in that
subset is the best fit for the ``dominant edges'' set.

In summary, we present here a model of weighted-network growth based
on a {\em weighted preferential attachment} principle \cite{P}: the
probability of picking an edge depends on the total weight of that
edge (and not simply on the number of times it has been
picked)\cite{KSBBHS,ZTZH}. We provide a theoretical framework, which
accounts for the empirical evidence that many real networks grow in a
heterogeneous way generating a subset of dominant edges that controls
a major share of the total weight of the network, while the weight of
other connections is negligible. Our approach is quite general and
flexible since it does not require a particular probability
distribution or functional form of the weights. Furthermore our model
produces in a natural way dense benchmark networks that can be used as
a reference or benchmark towards real dense networks. The mapping with
RRU has allowed us to introduce a statistical procedure for making
inference on the class of dominant links.  Thanks to the above
procedure, it is now possible to quantitatively test the convergence
to steady state in network dynamics, a problem often encountered in
assessing the significance of observations in complex networks.

Authors acknowledge support from CNR, PNR project ``CRISIS Lab'' and
FET Open project 255987 FOC.


\begin{thebibliography}{99}

\bibitem{AB} R. Albert, A.L. Barab\'asi, 
{\em Review of Modern Physics}, {\bf 74} (2002), 47-97.

\bibitem{castri}
D. Garlaschelli, S. Battiston, M. Castri, V.D.P. Servedio, G. Caldarelli, 
{\em Physica A} {\bf 350}, 491 (2005)

\bibitem{W} M. Kitsak, M. Riccaboni, S. Havlin, F. Pammolli,
H.E. Stanley, 
{\em Physical Review E}, {\bf 81}, 036117 (2010).

\bibitem{T} J. Tinbergen, Shaping the World Economy:
Suggestions for a International Economic Policy, New York: The
Twentieth Century Fund, (1962).

\bibitem{Fcube} R. Pastor-Satorras, A. Vespignani {\em Evolution and
  Structure of the Internet}, Cambridge Unviersity Press (2004).

\bibitem{newmanbook} 
M.E.J. Newman, {\em Networks: an introduction} Oxford University Press (2010).

\bibitem{batty} 
C. Roth, S.M. Kang, M. Batty, M. Barth\'elemy,  
{\em PLoS ONE} {\bf 6},  e15923. (2011).

\bibitem{mybook} G. Caldarelli,{\em Scale-Free Networks}, Oxford
  University Press (2007).

\bibitem{AMS} G. Aletti, C. May, P. Secchi, 
{\em Advances in Applied Probability}, 41 (2009), 829-844.

\bibitem{BCPR} P. Berti, I. Crimaldi, L. Pratelli, P. Rigo, 
{\em Journal of Applied Probabilities}, {\bf 48}, 527-546 (2011).

\bibitem{BCPRdom} P. Berti, I. Crimaldi, L. Pratelli, P. Rigo, 
{\em Stochastic Processes and their Applications}, {\bf120}, 1473-1491 (2010).

\bibitem{C} I. Crimaldi, 
{\em International Mathemaical Forum}, {\bf 23} , 1139-1156 (2009).

\bibitem{MF} C. May, N. Flournoy, 
{\em Annals of Statistics}, {\bf 37}, 1058-1078 (2009).

\bibitem{SB} A. Serrano, M. Bogu\~n\'a, 
{\em Physical Review E}, {\bf 68}, 015101(R) (2003).

\bibitem{GL-2004} D. Garlaschelli, M. Loffredo, 
{\em Physical Review Letters}, {\bf 93}, 188701 (2004).

\bibitem{HMR} E. Helpman, M. Melitz, Y. Rubinstein, 
{\em NBER working paper series}, 12927 (2007).

\bibitem{BMSKM-2008} K. Bhattacharya, G. Mukherjee, J. Saram\"aki,
K. Kaski, S.S. Manna, 
{\em Journal of Statistical Mechanics}, P02002 (2008).

\bibitem{FRS-2009} G. Fagiolo, J. Reyes, S. Schiavo, 
{\em Physical Review E},  {\bf 79}, 036115 (2009).

\bibitem{RS} M. Riccaboni, S. Schiavo, 
{\em New Journal of Physics}, {\bf 12}, 023003 (2010).

\bibitem{GCC} D. Garlaschelli, A. Capocci, G. Caldarelli,
{\em Nature Physics}, {\bf 3} , 813-817 (2007).

\bibitem{GL-2005} D. Garlaschelli, M. Loffredo, 
{\em Physica A}, {\bf 355}, 138-144 (2005).

\bibitem{H} K. Head, Gravity for beginners, (2003), available at\\ 
{http://economics.ca/keith/gravity.pdf}

\bibitem{colizza}
V. Colizza, A. Flammini, M. A. Serrano and A. Vespignani,
{\em Nature Physics} {\bf 2}, 110 - 115 (2006).

\bibitem{gin} V. Zlati\'c, G. Bianconi, A. D\'{\i}az-Guilera,
  D. Garlaschelli, F. Rao, G. Caldarelli, 
  {\em European Physical  Journal B} {\bf 67}, 271-275 (2009).

\bibitem{PNAS} A. Barrat, M. Barth\'elemy, A. Vespignani,
{\em Physical Review Letters}, {\bf 92}, 228701 (2004).

\bibitem{COMTRADE} United Nations Commodity Trade Statistics Database
http://comtrade.un.org/.




\bibitem{P} R. Pemantle, 
A survey of random processes with reinforcement, 
{\em Probability Surveys}  {\bf 4} , 1-79 (2007).


\bibitem{KSBBHS} T. Kalisky, S. Sreenivasan, L.A. Braunstein,
S.V. Buldyrev, S. Havlin, H.E. Stanley, 
{\em Physical Review E}, {\bf73}, 025103(R) (2006).

\bibitem{ZTZH} D. Zheng, S. Trimper, B. Zheng, P. Hui, 
{\em Physical Review E}, {\bf 67}, 040102 (2003).

\end{thebibliography}
\end{document}